\documentclass[11pt]{article}
\usepackage{moriond_plionis,epsfig}

\newcommand{\mincir}{\raise
-3.truept\hbox{\rlap{\hbox{$\sim$}}\raise4.truept\hbox{$<$}\ }}
\newcommand{\magcir}{\raise
-3.truept\hbox{\rlap{\hbox{$\sim$}}\raise4.truept\hbox{$>$}\ }}

\def\be{\begin{equation}}
\def\ee{\end{equation}}
\def\bea{\begin{eqnarray}}
\def\eea{\end{eqnarray}}

\begin{document}
\vspace*{4cm}
\title{HIGH-Z XRAY AGN CLUSTERING AND COSMOLOGICAL IMPLICATIONS}

\author{ Manolis Plionis}

\address{Institute of Astronomy \& Astrophysics, 
National Observatory of Athens, Greece \& \\ INAOE, Puebla, Mexico}

\maketitle\abstracts{
I review recent results of the high-redshift X-ray
selected AGN clustering, based on the XMM/2dF survey. Using the 
luminosity-dependent density evolution luminosity function we find that
the spatial clustering lengths, derived using Limber's inversion
equation, are $\sim 16$ and 19 $h^{-1}$ Mpc respectively (for the
comoving clustering evolution model) while the median redshifts of
the soft and hard X-ray sources are ${\bar z}\sim 1.2$ and 0.75, respectively.
Within the framework of flat
cosmological models we find that these results support a model 
with $\Omega_m \simeq 0.26$, $\sigma_8\simeq 0.75$, w$\simeq -0.9$ 
(in excellent agreement with the 3 year WMAP results). 
We also find the present day bias of X-ray AGNs to be $b_o\simeq 2$.}

Active Galactic Nuclei (AGN) can be detected out to high redshifts and
thus their clustering properties can provide
information on the large scale structure, the underlying matter
distribution and the evolution with redshift of the AGN phenomenon.
From the optical 2QZ and SDSS surveys 
it appears that the QSO clustering properties are comparable
to those of local galaxies (eg. Croom et al. 2002; 2005; Porciani,
Magliocchetti \& Norberg 2004; Wake et al. 2004), while there is 
also evidence for the
comoving clustering evolution model of active galaxies (see also
Kundi\'c 1997). 
Optically selected AGN catalogues however, miss large 
numbers of dusty systems and therefore, provide a biased census of the
AGN phenomenon. X-ray surveys, are least affected by dust providing an
efficient tool for compiling uncensored AGN samples over a 
wide redshift range. From the
cosmological point of view an interesting question that remains to be
addressed is how the high-$z$ X-ray selected  AGNs trace the underlying mass
distribution and whether there are any differences with optically
selected samples. 
 
Early studies of the X-ray AGN clustering,
using {\it Einstein} and {\it ROSAT} data, produced 
contradictory results with other studies finding significant
clustering while others not (Boyle \& Mo 1993; Vikhlinin \& Forman
1995; Carrera  et al. 1998; Akylas, Georgantopoulos \& Plionis 2000; 
Mullis et al. 2004).

Recently, there has been an effort to address this confusing issue and
determine the clustering
properties of both soft and hard X-ray selected AGNs, based on the 
new XMM and Chandra
missions (eg. Yang et al. 2003, 2006; Basilakos
et al 2004, 2005; Gilli, Daddi, Zamorani 2005; Puccetti et al. 2006; 
Gandhi et al. 2006). Most of these studies find a large correlation
length for the high-z X-ray AGNs, with hard sources having an even
larger correlation length than soft sources, 
with $r_{\circ}\sim 17-19 \;h^{-1}$ Mpc -
(eg. Basilakos et al. 2004; Puccetti et al. 2006). 

Here I review our recent results based on the 2 deg$^{2}$ XMM/2dF
survey, which exploits the high sensitivity and  the large
field-of-view of the XMM-{\it Newton} observatory.

\section{The XMM/2df survey: $\log N-\log S$ and w$(\theta)$}
The XMM-{\it Newton}/2dF survey is a shallow (2-10\,ksec per  pointing) 
survey comprising of 18 XMM-{\it
Newton} pointings equally split between a Northern and Southern
Galactic region
near the corresponding poles\footnote{{\em North:} RA(J2000)=$13^{\rm
    h} 41^{\rm m}$; Dec.(J2000)=$00^{\circ} 00^{'}$]
and {\em South:}  RA(J2000)=$00^{\rm h} 57^{\rm m}$,
Dec.(J2000)=$-28^{\circ} 00^{'}$}. 
Due to elevated particle background we analysed 
a total of 13 pointings. A full description of
the data reduction, source detection and flux estimation are presented
in Georgakakis et al. (2003, 2004). 


We will present results using the soft (0.5-2\,keV), hard (2-8\, keV)
and total  (0.5-8\,keV) band
catalogues of the XMM-{\it Newton}/2dF survey. We only consider
sources at off-axis angles $<13.5$\,arcmin. These samples comprise
of 432, 171 and 462 sources respectively above the 
$5\sigma$  detection threshold. The limiting fluxes are 
$f_X(\rm 0.5-2)=2.7\times 10^{-15} \, erg \, \, s^{-1} cm^{-2}$,
$f_X(\rm 2-8)= 10^{-14} \, erg \, \, s^{-1} cm^{-2}$ 
and $f_X(\rm
0.5-8)=6.0\times 10^{-15} \, erg \, \, s^{-1} cm^{-2}$. 

We derive the source $\log N- \log S$ after constructing
sensitivity maps in order to estimate the
area of the survey accessible
to point sources above a given flux limit (see Basilakos et al. 2004, 2005).
In the 0.5-2\,keV band there is
good agreement between our results and the Baldi et al. (2002) double 
power-law best fit to the number counts. The  Manners et al. (2003)
best fit is derived for sources in the flux range $f(\rm 0.5-8\,keV)
\rm = 10^{-15} - 8 \times 10^{-14} \, erg \, s^{-1} \,
cm^{-2}$. Although our $dN/dS$ is in good agreement with their  
results in the above flux range, at brighter fluxes the surface 
density of X-ray sources is lower than the extrapolated Manners et
al. (2003) relation. This suggests that a double power-law is required to
fit the 0.5-8\,keV $dN / dS$ over the flux range $\rm 10^{-15} -
10^{-12} \, erg \, s^{-1} \, cm^{-2}$, which indeed provides an
excellent fit to our $\log N- \log S$ (see Basilakos et al. 2005).

We then calculate the angular correlation function using the 
estimator: w$(\theta)=f N_{DD}/N_{DR} - 1$, 
of which the uncertainty is: $\sigma_{w}=\sqrt{(1+{\rm w}(\theta))/N_{DR}}$,
where $N_{DD}$ and $N_{DR}$ are the number of data-data and
data-random pairs, respectively, in the interval
$[\theta-\Delta \theta,\theta+\Delta \theta]$. 
The normalization factor is $f = 2 N_R /(N_D-1)$, with
$N_D$ and $N_R$ the total number of data and random points,
respectively. 
For each XMM pointing we produce 100 Monte Carlo random 
catalogues having the same number of points as the real data 
 which also account for the sensitivity variations across 
the surveyed area (see
section 2). 
Furthermore since the flux threshold for source detection depends on
the off-axis angle from the center of each of the XMM-{\it Newton} 
pointing, the sensitivity maps are used to discard random points in
less sensitive areas. This is
accomplished by assigning a flux to each random point using the
source $\log N- \log S$. If
that flux is less than 5 times the local {\em rms}  noise at the
position of the random point (assuming Poisson statistics for the
background) this is excluded from the random data-set. We have
verified that our random simulations reproduce 
both the off-axis sensitivity of the detector as well as 
the individual field $\log N - \log S$.
Using the methods described above we estimate w$(\theta)$ in
logarithmic intervals with $\delta \log \theta\simeq 0.05$. 
For all three samples we estimate w$(\theta<150^{''})$ and 
find a statistically significant signal (see Table 1) at a $\magcir 3 \sigma$
confidence level (Poisson statistics). 
We then use a standard $\chi^{2}$ minimization procedure to
fit the measured  correlation function assuming a power-law
form: w$(\theta)= (\theta_{\circ} /  \theta) ^ {\gamma-1}$ and fixing
$\gamma$ to 1.8. 
Note that (a) the fitting is performed for
angular separations in the range 40--1000\,arcsec, and (b)
our results are insensitive to both the upper cutoff limit in
$\theta$ and the angular binning (for more than 10 bins)
used to estimate  w$(\theta)$. The resulting raw values of $\theta_o$
are corrected for the integral constraint and the amplification bias
(see Vikhlinin \& Forman 1995 and Basilakos et al. 2005), although
such corrections are quite small. The final results are presented in Table 1.

\tabcolsep 9pt
\begin{table}
\tabcolsep 14pt
\begin{tabular}{cccccc} 
\hline
X-ray band& No. of sources& $\theta_{\circ}$(arcsec)& $\chi^{2}/{\rm
  dof}$ & $P_{\chi^{2}}$ & w$(\theta <150^{''})$ \\ \hline 
 0.5-8\,keV  &  462 &  $10.8 \pm 1.7$& 1.50& 0.10& $0.114\pm 0.037$\\
 0.5-2\,keV  &  432 &  $10.4 \pm 1.9$& 1.10& 0.35& $0.105\pm 0.035$\\
 2-8\,keV    &  177 &  $28 \pm 9$    & 0.88 & 0.57 & 0.128$\pm 0.080$ \\
\hline
\end{tabular}
\caption{Angular correlation function analysis results with their 
$1\sigma$ ($\Delta \chi^{2}=1.00$) uncertainties. The fits are
  produced after imposing $\gamma=1.8$.} 
\end{table}

\begin{figure}
\mbox{\epsfxsize=16cm \epsffile{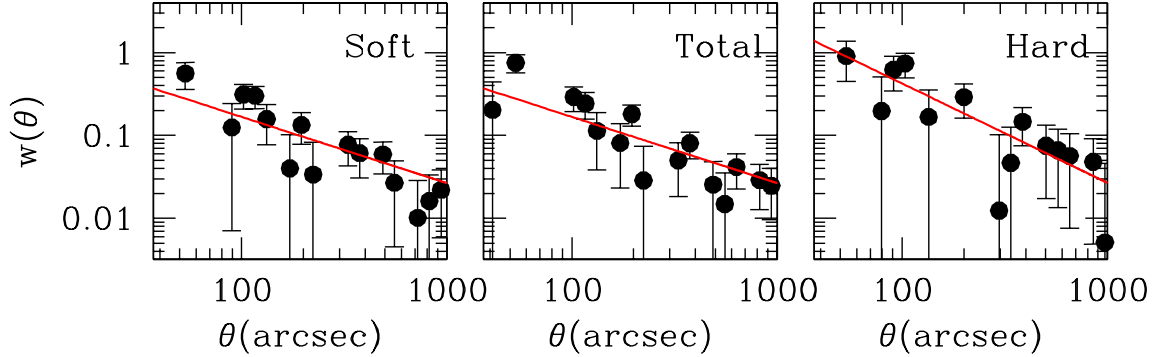}}
\caption{The XMM/2dF survey two-point angular correlation function 
for the 3 bands considered.}
\end{figure}

\section{The spatial correlation length of the XMM/2dF soft X-ray sources}
The spatial correlation function can be modeled as 
(eg. de Zotti et al. 1990):
\begin{equation}\label{eq:dezotti}
\xi(r,z)=(r/r_{\circ})^{-\gamma}\times (1+z)^{-(3+\epsilon)}\;\;,
\end{equation} 
where $\epsilon$ parametrizes the type of clustering evolution.
For $\epsilon=\gamma-3$ (ie., $\epsilon=-1.2$ for $\gamma=1.8$), 
the clustering is constant in comoving coordinates (comoving
clustering), a model which appears to be
 appropriate for active galaxies (eg. Kundi\'c 1997; Croom et al. 2005).

In order to invert the angular correlation function to three dimensions
we utilize Limber's integral equation (eg. Peebles 1993). 
For a spatially flat Universe, Limber equation can be written as:
\be 
{\rm w}(\theta)=2\frac{\int_{0}^{\infty} \int_{0}^{\infty} x^{4} 
\phi^{2}(x) \xi(r,z) {\rm d}x {\rm d}u}
{[\int_{0}^{\infty} x^{2} \phi(x){\rm d}x]^{2}} \;\; , 
\ee 
where $\phi(x)$ is the distance selection function (the probability 
that a source at a distance $x$ is detected in the survey), 
$x$ is the proper distance related to the redshift through (see Peebles 1993): 
\be
x(z)=\frac{c}{H_{\circ}} \int_{0}^{z} \frac{{\rm d}t}{E(t)}\;\; \;\; {\rm
  with}  \;\;\; 
E(z)=[\Omega_{\rm m}(1+z)^{3}+\Omega_{\Lambda}]^{1/2}\;\;,
\ee
The selection function, $\phi(x)$, is related to the
number of objects in the given 
survey with a solid angle $\Omega_{s}$ and within 
the shell $(z,z+{\rm d}z)$, by:
\be
\frac{{\rm d}N}{{\rm d}z}=\Omega_{s}
x^{2}\phi(x)\left(\frac{c}{H_{\circ}}\right) E^{-1}(z)\;\;.
\ee
Since we do not have complete redshift information for our sources 
we estimate ${\rm d}N/{\rm d}z$ using the X-ray source luminosity function
and the specific flux-limit of our samples, via the relation:  
$\phi(x)=\int_{L_{\rm min}(z)}^{\infty} \Phi(L_{x},z) {\rm d}L$,
where $\Phi(L_{x},z)$ is the 
luminosity dependent density evolution luminosity (LDDE)
function. For the soft-band we use that 
of Miyaji, Hasinger \& Schmidt (2000) while for the hard-band that
of Ueda et al (2003). In
Fig. 2 we present the expected redshift distributions of the
soft and hard X-ray sources together with the histogram of some
limited spectroscopic and photo-z data (see caption for details).
 The LDDE model predicts a redshift
distribution with a
median redshift of $\bar{z} \simeq 1.2$ and 0.75 for the soft and hard
sources respectively.
Finally, the expression for w$(\theta)$ satisfies the form:
\begin{equation}\label{eq:angu}
{\rm w}(\theta)=2\frac{H_{\circ}}{c} \int_{0}^{\infty} 
\left(\frac{1}{N}\frac{{\rm d}N}{{\rm d}z} \right)^{2}E(z){\rm d}z 
\int_{0}^{\infty} \xi(r,z) {\rm d}u
\end{equation} 
Note that, the physical separation between two sources, 
separated by an angle $\theta$ considering 
the small angle approximation, is given by:
$r \simeq (1+z)^{-1} \left( u^{2}+x^{2}\theta^{2} \right)^{1/2}$.


\begin{figure}\label{fig:reds}
\mbox{\epsfxsize=15cm \epsffile{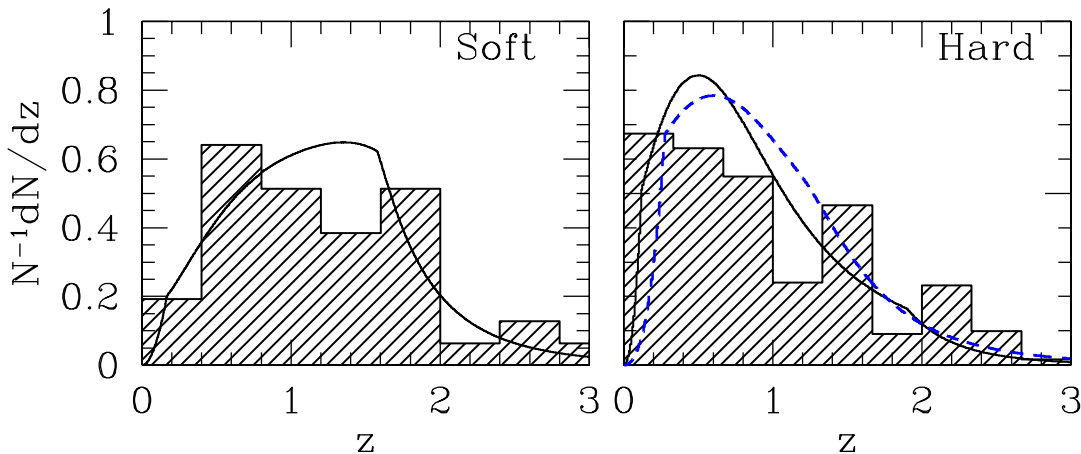}}
\caption{{\em Left Panel:} The expected redshift distribution
(continuous line) by the
Miyaji et al (2000) LDDE soft-band luminosity
function. {\em Right Panel:} The corresponding redshift distribution
for the Ueda et al. (2003) LDDE hard-band luminosity function (the
dashed line is based on the La Franca et al. 2005 luminosity function).
The histogram in the left panel corresponds 
to the distribution of the Schmidt et al (1998) X-ray 
sources of the ROSAT Lochman Deep Field (albeit having a flux limit
slightly lower than of our survey), while that of the right panel
on limited spectroscopic and photo-z data of our XMM/2dF survey.}
\end{figure}


%

Then the inversion of eq. (5), using the LDDE luminosity evolution model,
$\epsilon=-1.2$ and the concordance cosmological model, provides a spatial
correlation length of $r_{\circ}\simeq 16.4\pm 1.3 \;h^{-1}$ Mpc
and $\simeq 19\pm 1.3 \;h^{-1}$ Mpc, for the
soft and hard bands, respectively. 
These results are in very good agreement with
a recent XMM based study of the ELAIS-S1 field by Puccetti et
al. (2006)
and comparable to those of Extremely 
Red Objects (EROs), of luminous radio  sources (Roche, Dunlop \& 
Almaini 2003; Overzier et al. 2003; R\"{o}ttgering et al. 2003) and of
bright distant red galaxies (Foucaud et al. 2006).

Our $r_{\circ}$ values, however, are significantly
larger than those 
derived from optical AGN surveys (which trace mostly the unobscured
component): $r_{\circ} \simeq 5.4 - 8.6 \; h^{-1}$ Mpc
(eg. Croom \& Shanks 1996; La Franca et al. 1998; 
Croom et al. 2002; Grazian et al. 2004; Porciani et al. 2004; Wake et
al. 2004). 
We can push our inverted $r_{\circ}$ values to approximate closely 
the optical AGN results only if we use the constant in physical coordinates
clustering evolution model 
($\epsilon=-3$), in which case we obtain $r_{\circ} \simeq 7.5\pm 0.6 
\;h^{-1}$ and $\simeq 13.5\pm 3 
\;h^{-1}$ Mpc (for the soft and hard bands respectively). 

\section{Cosmological Constraints}
It is well known (Kaiser 1984) that according to
linear biasing the correlation function of the mass-tracer 
($\xi_{\rm obj}$) and dark-matter one ($\xi_{\rm DM}$), are related by:
\be
\label{eq:spat}
\xi_{\rm obj}(r,z)=b^{2}(z) \xi_{\rm DM}(r,z) \;\;, 
\ee
where $b(z)$ is the bias evolution function. In this study 
we use the bias model
of Basilakos \& Plionis 2001; 2003) 
which is based on linear perturbation theory and the 
Friedmann-Lemaitre solutions of the cosmological
field equations. 
We quantify the underlying matter distribution clustering 
by presenting the spatial correlation function of the mass 
$\xi_{\rm DM}(r,z)$ 
as the Fourier transform of the 
spatial power spectrum $P(k)$:
\be
\label{eq:spat1}
\xi_{\rm DM}(r,z)=\frac{(1+z)^{-(3+\epsilon)}}{2\pi^{2}}
\int_{0}^{\infty} k^{2}P(k) 
\frac{{\rm sin}(kr)}{kr}{\rm d}k \;\;,
\ee
where $k$ is the comoving wavenumber and
$\epsilon=-1.2$, according to the constant in comoving coordinates
clustering evolution model.
As for the power spectrum, we consider that of CDM models, 
where $P(k)=P_{0} k^{n}T^{2}(k)$ with
scale-invariant ($n=1$) primeval inflationary fluctuations. 
In particular, we use the transfer function parameterization as in
Bardeen et al. (1986), with the corrections given approximately
by Sugiyama (1995). The
normalization of the power spectrum is given by: 
\be 
P_{0}=2\pi^{2} \sigma_{8}^{2} \left[ \int_{0}^{\infty} T^{2}(k)
  k^{n+2} W^{2}(kR){\rm d}k \right]^{-1} \;\;. 
\ee
where $\sigma_{8}$ is the rms mass fluctuation
on $R=8h^{-1}$Mpc scales and $W(kR)$ is the window function.
Note that we also use the
non-linear corrections introduced by Peacock \& Dodds (1994).  

We have chosen to use either the standard normalization
given by: $\sigma_8 \simeq 
0.5 \Omega_{\rm m}^{-\gamma}$ with $\gamma \simeq 0.21-0.22 {\rm w} +0.33
\Omega_{\rm m}$ (Wang \& Steinhardt 1998), 
or to leave $\sigma_8$ a free parameter to be fitted by
our analysis.

\subsection{X-ray AGN Clustering likelihood}
It has been shown that the
application of the correlation function analysis on samples of
high redshift galaxies can be used as a useful tool 
for cosmological studies (eg. Matsubara 2004).
In what follows we review a similar analysis, presented in Basilakos
\& Plionis (2005, 2006), utilizing a $\chi^{2}$ 
likelihood procedure to compare the measured 
XMM soft source angular correlation function  
with the prediction of different spatially flat cosmological models.
In particular, we define the likelihood estimator
as:
${\cal L}^{\rm AGN}({\bf c})\propto {\rm exp}[-\chi^{2}_{\rm AGN}({\bf c})/2]$
with:
\be
\chi^{2}_{\rm AGN}({\bf c})=\sum_{i=1}^{n} \left[ \frac{ w_{\rm th}
(\theta_{i},{\bf c})-w_{\rm obs}(\theta_{i}) }
{\sigma_{i}} \right]^{2} \;\;.
\ee 
where ${\bf c}$ is a vector containing the cosmological 
parameters that we want to fit and $\sigma_{i}$ the observed angular 
correlation function uncertainty. 
We assume a flat ($\Omega_{\rm tot}=1$) cosmology 
with primordial adiabatic fluctuations and baryonic
density of $\Omega_{\rm b} h^{2}\simeq 0.022$ 
(eg. Kirkman et al. 2003; Spergel et al. 2006). 
In this case the corresponding vector
is ${\bf c}\equiv (\Omega_{\rm m},{\rm w},\sigma_8,h,b_{\circ})$
and we sample the various parameters as follows:
the matter density $\Omega_{\rm m} \in [0.01,1]$ in steps of
0.01; the equation of state parameter $w\in [-3,-0.35]$ in steps
of 0.05, the power spectrum normalization $\sigma_8 \in [0.4,1.4]$ in
steps of  0.02, the dimensionless Hubble constant $h \in [0.5,0.9]$ in steps
of 0.02 and the X-ray sources bias at the present time
$b_{\circ} \in [0.5,4]$ in steps of 0.05.
Note that in order to investigate possible equations of state,
we have allowed the parameter $w$ to take values below -1. Such models 
correspond to the so called {\em phantom} cosmologies (eg. Caldwell 2002).

The resulting best fit parameters for the $\epsilon=-1.2$ clustering
evolution model are presented in Table 2. In the first two rows we
present results based on the traditional Wang \& Steinhardt (1998)
$\sigma_8$ normalization. Note that our estimate of the Hubble
parameter $h$ (left panel in Fig. 3)
is in very good agreement with those derived ($h=0.72\pm
0.07)$ by the HST key project (Freeman et al. 2001).
In the last two rows of Table 2 we leave 
$\sigma_8$ free but fix the Hubble constant to $h=0.72$.  
In Fig.3 we present the 1$\sigma$, 2$\sigma$ and $3\sigma$
confidence levels in the various parameter planes,
marginalizing over the rest of the parameters.
We find that $w$ is
degenerate with respect to both $h$ and the bias at the present time.
\begin{figure}
\mbox{\epsfxsize=16.3cm \epsffile{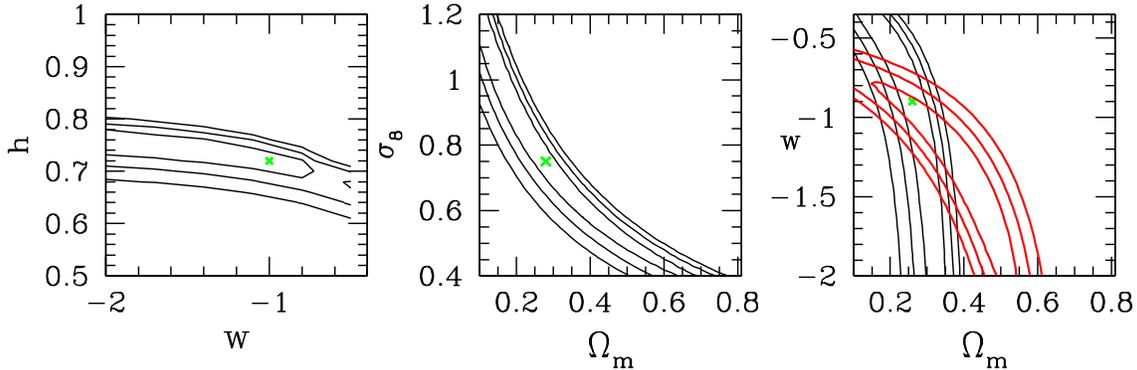}}
\caption{Likelihood contours in the $({\rm w},h)$ plane (left panel),
the $(\sigma_8,\Omega_{\rm  m})$ plane (central panel) 
and the $(w,\Omega_{\rm m})$ plane (right panel). The
contours are 
plotted where $-2{\rm ln}{\cal L}/{\cal L}_{\rm max}$ is equal
to 2.30, 6.16 and 11.83, corresponding 
to 1$\sigma$, 2$\sigma$ and 3$\sigma$ confidence level.
In the $(\Omega_{\rm m},w)$
plane we plot as thick lines the likelihood contours derived from the
SNIa Hubble relation.}
\end{figure}

When we leave free the $\sigma_8$ parameter, our fit (central panel of
Fig. 3) provides a value which is in excellent agreement with that derived 
by the recent 3-years WMAP results (Spergel et al. 2006)
\footnote{Hereafter, when we marginalize over the 
equation of state parameter we will use $w=-1$.}. 
Therefore, 
allowing for the first time values $w<-1$ 
(Phantom models) we can derive a 
$(\Omega_{\rm m},\sigma_{8})$ relation, a good fit of which is
provided by :
\be 
\sigma_{8}=0.34 (\pm 0.01) \; \Omega_{\rm m}^{-\gamma(\Omega_{\rm m},w)}
\ee
with $\gamma(\Omega_{\rm m},w)=0.22 (\pm 0.04)-0.40 (\pm 0.05)w-0.052 (\pm
0.040)\Omega_{\rm m}$.

Note that eq. (10) produces $\sigma_{8}$ values which are
significantly smaller than the usual cluster normalization (Wang \&
Steinhardt 1998) but are 
in good agreement with the 3-years WMAP results;  
for example for $w\simeq -1$ and $\Omega_{\rm m}\simeq 0.28$
we get $\sigma_{8}\simeq 0.73\pm 0.03$. 

Inspecting the thin contours in the right panel of Fig. 3 
it becomes evident that $w$ is
degenerate, within the $1\sigma$ uncertainty,
with respect to $\Omega_{\rm m}$.
Therefore, in order to put further constraints on 
$w$ we additionally use a sample of 172 supernovae SNIa (see Tonry et
al. 2003).

\begin{table}
\tabcolsep 9pt
\begin{tabular}{cccccc} 
\hline
Data& $\Omega_{\rm m}$& $\sigma_8$  &$w$& $h$& $b_{\circ}$ \\ \hline 
{\rm XMM} &  $0.31^{+0.16}_{-0.08}$ & 0.93  & uncons. ($w=-1$) &
$0.72^{+0.02}_{-0.18}$&  $2.30^{+0.70}_{-0.20}$ \\
{\rm XMM}/{\rm SNIa}& $0.28\pm 0.02$ &0.95  & $-1.05^{+0.10}_{-0.20}$&
$0.72$& $2.30$ \\
{\rm XMM} &  $0.28 \pm 0.03$ & $0.75 \pm 0.03$  & uncons.($w=-1$) & 0.72&  $2.0^{+0.20}_{-0.25}$\\
{\rm XMM}/{\rm SNIa}& $0.26\pm 0.04$ &0.75  &
$-0.9^{+0.10}_{-0.05}$&$0.72$ & $2.0$\\ \hline
\end{tabular}
\caption[]{Cosmological parameters from the likelihood analysis.
Errors of the fitted parameters 
represent $1\sigma$ uncertainties. Note that for the joined
analysis (2nd and 4th rows) the corresponding results are marginalized over 
the parameters that do not have errorbars, for which
we use the values indicated.}
\end{table}
 
\subsection{The AGN+SNIa likelihoods}
We combine the X-ray AGN clustering properties 
with the SNIa data by performing a joined likelihood analysis and
marginalizing the X-ray clustering results 
over $\sigma_{8}$, $h$ and $b_{0}$. The vector ${\bf c}$ now becomes: 
${\bf c}\equiv (\Omega_{\rm m}, w)$. The SNIa 
likelihood function can be written as: 
${\cal L}^{\rm SNIa}({\bf c})\propto 
{\rm exp}[-\chi^{2}_{\rm SNIa}({\bf c})/2]$, 
with:
\be
\chi^{2}_{\rm SNIa}({\bf c})=\sum_{i=1}^{172} \left[ \frac{ {\rm log}
    D^{\rm th}_{\rm L}
(z_{i},{\bf c})-{\rm log}D^{\rm obs}_{\rm L}(z_{i}) }
{\sigma_{i}} \right]^{2} \;\;,
\ee 
where $D_{\rm L}(z)$ is the dimensionless luminosity
distance, $D_{\rm L}(z)=H_{\circ}(1+z)x(z)$
and $z_{i}$ is the observed redshift. 
The results are shown as thick lines in Fig. 3 and represent 
the $1\sigma$, $2\sigma$, and $3\sigma$,  
confidence levels.
The joint likelihood function 
(${\cal L}^{\rm joint}(\Omega_{\rm m}, w)={\cal L}^{\rm AGN} \times 
{\cal L}^{\rm SNIa}$) peaks at: 
$\Omega_{\rm m}=0.26 \pm 0.04$ with $w=-0.90^{+0.1}_{-0.05}$.
Using eq. (10) we find that the normalization
of the power spectrum that corresponds to these cosmological
parameters is
$\sigma_{8}\simeq 0.73$. It should be pointed out that our 
results are 
in excellent agreement with the recent 3-year WMAP results of
Spergel et al. (2006).

Other recent analyzes, utilizing different combinations of data,
seem to agree with our results.
For example, Sanchez et al. (2006) used the WMAP (1-year) CMB
anisotropies in combination with the 2dFGRS power spectrum and found 
$\Omega_{\rm m}\simeq 0.24$ and $w\simeq -0.85$, while Wang \&
Mukherjee (2006) utilizing the 3-years WMAP data together 
with SNIa and galaxy clustering results found $w\simeq
-0.9$. 

\section*{Acknowledgments}
This presentation is based on the results of a
Hellenic scientific collaboration project, entitled: 
{\em 'X-ray Astrophysics with ESA's mission XMM'}, jointly funded by the European Union
and the Greek Government in the framework of the program 'Promotion
of Excellence in Technological Development and Research'. The colleagues
involved in the research presented here are: Spyros Basilakos,
Ioannis Georgantopoulos, Antonis
Georgakakis and myself. I especially thank Spyros Basilakos for useful
discussions.

\end{document}